# Maximizing Spin-Orbit Torque Generated by the Spin Hall Effect of Pt

Lijun Zhu[1,2*], Daniel C. Ralph[1,3], Robert A Buhrman[1]

*1. Cornell University, Ithaca, New York 14850, USA*
*2. State Key Laboratory of Superlattices and Microstructures, Institute of Semiconductors, Chinese Academy of Sciences, P.O. Box 912, Beijing 100083, China*
*3. Kavli Institute at Cornell, Ithaca, New York 14850, USA*
*[*lz442@cornell.edu](mailto:lz442@cornell.edu)*

Efficient generation of spin-orbit torques is central for the exciting field of spin-orbitronics. Platinum, the archetypal spin Hall material, has the potential to be an outstanding provider for spin-orbit torques due to its giant spin Hall conductivity, low resistivity, high stabilities, and the ability to be compatible with CMOS circuits. However, pure clean-limit Pt with low resistivity still provides a low damping-like spin-orbit torque efficiency, which limits its practical applications. The efficiency of spin-orbit torque in Pt-based magnetic heterostructures can be improved considerably by increasing the spin Hall ratio of Pt and the spin transmissivity of the interfaces. Here we review recent advances in understanding the physics of spin current generation, interfacial spin transport, and the metrology of spin-orbit torques, and summarize progress towards the goal of Pt-based spin-orbit torque memories and logic that are fast, efficient, reliable, scalable, and non-volatile.

## TABLE OF CONTENTS

## I. INTRODUCTION

Efficient manipulation of magnetization in magnetic nanostructures is essential for spintronic devices. Since the discovery that an in-plane charge current in certain heavy metal (HM) thin films can be utilized to effectively manipulate the magnetization state of an adjacent ferromagnetic layer,[1–6] spin-orbit torques (SOT) have become a powerful and versatile tool to manipulate and excite any type of magnetic materials, ranging from metals to semiconductors and insulators, in ferromagnetic, ferrimagnetic, and antiferromagnetic configurations. Strong SOTs can enable fast, efficient, reliable, and scalable magnetic memories and logic that utilize magnetic tunnel junctions (MTJs)[5,7-9], magnetic domain walls,[10,11] or magnetic skyrmions[12,13] as the information storage elements. SOTs are exerted when angular momentum is transferred from a spin current or spin accumulation generated by a variety of possible spin-orbit-coupling (SOC) effects, e.g. the bulk spin Hall effect (SHE),[14-17] topological surface states,[18,19] interfacial SOC effects,[20-24] orbit-spin conversion,[25] the anomalous Hall effect,[26] the planar Hall effect,[27,28] the magnetic SHE,[29-31] etc.

As shown in Fig. 1(a), a spin current with spin polarization $\sigma$ can exert two types of SOTs on a magnetization $M$ of a thin-film ferromagnet (FM), i.e., the damping-like torque [$\tau_{DL} \sim M \times (M \times \sigma)$] due to the absorption of the spin current component transverse to $M$ and field-like torque [$\tau_{FL} \sim (M \times \sigma)$] due to reflection of the spin current with some spin rotation. The same physics can be expressed in terms of effective SOT fields: a damping-like effective SOT field ($H_{DL}$) parallel to $M \times \sigma$ and a field-like effective SOT field ($H_{FL}$) parallel to $\sigma$. The damping-like SOT is technologically more important because it can excite dynamics and switching of magnetization (even for low currents for which $H_{DL}$ is much less than the anisotropy field of a perpendicular magnetization). Field-like SOTs by themselves can destabilize magnets only if $H_{FL}$ is greater than the anisotropy field, but they can still strongly affect the dynamics in combination with a damping-like SOT.[32]

When a bulk SHE is the predominant source of the spin current responsible for the damping-like SOT, as is usually the case for magnetic heterostructures containing thin-film heavy metal (HM),[3-10,32-34] Bi-Sb,[35] Bi$_x$Te$_{1-x}$,[36] CoPt,[37] and Co-Ni-B[38] [see Fig. 1(b)], a damping-like SOT efficiency per applied electric field ($\xi_{DL}^E$) and a damping-like SOT efficiency per unit current density ($\xi_{DL}^j$) are given by

$$\xi_{DL}^E = (2e/\hbar)T_{int}\sigma_{SH}, \qquad (1)$$

$$\xi_{DL}^j = T_{int}\theta_{SH}. \qquad (2)$$

Here, $e$ is the elementary charge, $\hbar$ is the reduced Plank's constant, $T_{int}$ is the spin transparency of the magnetic interface (see detailed discussions in Section III), $\sigma_{SH}$ and $\theta_{SH}$ are the spin Hall conductivity and the spin Hall ratio (more precise than the "spin Hall angle" since it is really not an angle for high values, especially those approaching or greater than 1) of the spin Hall layer. The spin Hall ratio of a material is defined as

$$\theta_{SH} = (2e/\hbar)\, j_s/j_c, \qquad (3)$$

where $j_c$ is the density of the change current flow in the spin Hall layer and $j_s$ is the emitted spin current density from the spin Hall layer (which usually differs from the spin current density absorbed by the adjacent magnetic layer by $T_{int}$). The spin Hall conductivity is related to the spin Hall ratio by the electrical conductivity ($\sigma_{xx}$) of the spin Hall layer:

$$\sigma_{SH} = \theta_{SH}\, \sigma_{xx}. \qquad (4)$$

In comparison of different samples, $\xi_{DL}^j$ and $\xi_{DL}^E$ better quantify the efficiencies of damping-like SOT compared to $H_{DL}$ because the latter varies inversely with the saturation magnetization ($M_s$) and the layer thickness ($t_{FM}$) of the driven FM layer. To obtain a high $\xi_{DL}^j$ in magnetic heterostructure, it is necessary to achieve both a high $\theta_{SH}$ and a high $T_{int}$.

Since the first studies of spin Hall physics in metals beginning more than a decade ago,[16,17,39-45] platinum (Pt), the archetypal spin Hall material, has been central in generating and detecting pure spin currents and key in establishing most of the recent SOC phenomena, including the spin Seebeck effect,[46,47] spin-torque oscillators,[48,49] spin-torque switching of magnetization,[2,4,50,51] the spin Hall magnetoresistance,[52] SOT-driven chiral domain wall motion,[10,53] and ultrafast SOT-driven magnetic memories.[32] Despite the low $\xi_{DL}^j$ in the clean limit ($\xi_{DL}^j < 0.06$ at the resistivity $\rho_{xx}$ of 20 μΩ cm),[3] Pt still provides a particularly interesting base for the development of energy-efficient, high-endurance, and integration-friendly spin-current generators because its spin Hall conductivity is quite high compared to other heavy metals (e.g. Ta,[5] W,[6] and Pd[54]). Recent efforts have significantly advanced the understanding of the charge-spin conversion in Pt and spin transport phenomena through Pt interfaces, enhancing $\xi_{DL}^j$ considerably so that Pt-based alloys and heterostructures are now some of the most compelling sources of SOT for future technologies.

Since there have been several nice review papers discussing the general aspects of current-induced SOTs,[55-62] our purpose here is to provide a focused review of the very recent progress in understanding charge-spin conversion and interfacial spin transport in Pt-based HM/FM systems as well as the resulting successes in maximizing the SOT efficiencies and in optimizing Pt-based devices for SOT applications.

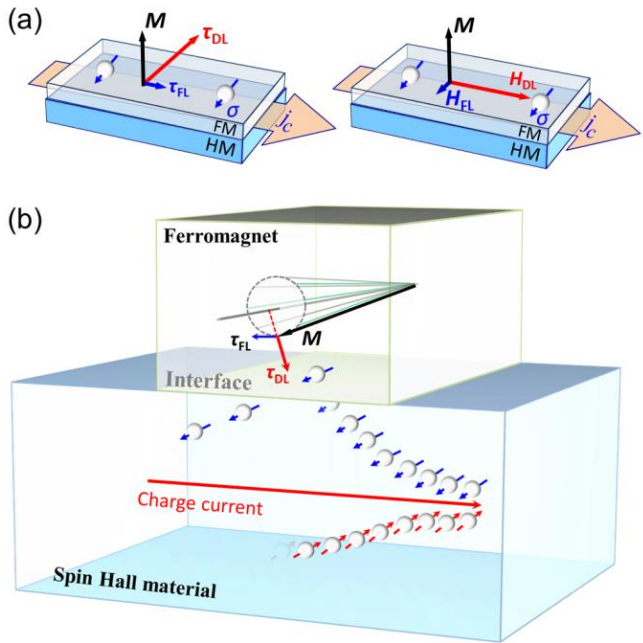

FIG. 1. (a) Directions of damping-like and field-like SOTs, $\tau_{DL}$ and $\tau_{FL}$, and their effective fields $H_{DL}$ and $H_{FL}$ on a perpendicular magnetization in a heavy metal/ferromagnet bilayer (HM/FM, $\theta_{SH} > 0$). (b) Schematic of SOT process in a spin Hall material/ferromagnet bilayer ($\theta_{SH} > 0$) with in-plane magnetic easy axis. A portion of the spin current generated by charge current flow in the spin Hall material diffuses through the interface and exerts dampinglike and fieldlike SOTs on the processional ferromagnet layer.

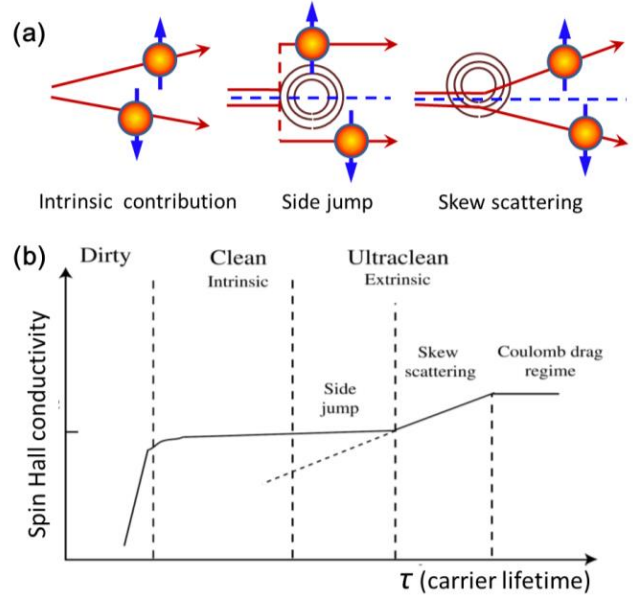

FIG. 2. (a) Schematic of the SHE mechanisms: intrinsic contribution, skew scattering, and side jump. (b) Evolution of the spin Hall conductivity of Pt with carrier lifetime. Reprinted with permission from Vignale, J. Supercond. Nov. Magn. 23, 3 (2010). Copyright 2009, Springer Science Business Media, LLC. The side jump and skew scattering can be important in the ultraclean limit but become negligible in the clean and dirty regimes, leaving the intrinsic contribution as the dominant mechanism in the clean and dirty regimes.

## II. SPIN HALL RATIO

### A. Spin Hall conductivity

While there is certainly on-going interest in searching for new spin torque mechanisms,[20-31,37,63] it is increasingly accepted that the dominant source of the SOTs in Pt-based HM/FM bilayers is the bulk SHE of Pt.[10,48,53,54] Understanding the mechanisms that produce the bulk SHE is therefore critical for optimizing the spin Hall ratio. As shown in Fig. 2(a), the bulk SHE of a HM can have three possible contributions: (i) the intrinsic contribution from the Berry curvature of the band structure and (ii) the extrinsic skew-scattering and (iii) side-jump contributions from spin-orbit-interaction-related impurity scattering.[56]

Given the relation $\theta_{SH} = \sigma_{SH}/\sigma_{xx}$ for the SHE, $\theta_{SH}$ should increase in any processes that increase the $\sigma_{SH}/\sigma_{xx}$ ratio. For intrinsic SHE, $\sigma_{SH}$ is independent of scattering potential in the clean limit but reduces with enhancing disorder level in the dirty limit.[64-66] Thus, for samples in the clean limit where the intrinsic SHE is dominant, $\theta_{SH}$ can be increased by decreasing $\sigma_{xx}$ until the $\sigma_{SH}/\sigma_{xx}$ ratio is maximized in the dirty limit. Analogous to the anomalous Hall effect in FM metals, $\sigma_{SH} \propto \sigma_{xx}^2/\sigma_{xx0}$ for the skew scattering and $\sigma_{SH} \propto \sigma_{xx}^2/\sigma_{xx0}^2$ for side jump when the impurity scattering and phonon scattering dominate the electron scattering of the metals.[67-69] Here $\sigma_{xx0}$ is the value of $\sigma_{xx}$ at low temperatures at which only impurity scattering is important for electron scattering. As a result, $\theta_{SH}$ for side jump may be enhanced by increasing $\sigma_{xx}/\sigma_{xx0}^2$. In the case of a "dirty metal" where $\sigma_{xx} \approx \sigma_{xx0}$, $\theta_{SH}$ for skew scattering is approximately independent of $\sigma_{xx}$ and $\rho_{xx}$.

Theoretically, the intrinsic contribution should dominate the spin Hall conductivity of Pt in the dirty and clean regimes where the carrier lifetime ($\tau$) is short, whereas the extrinsic contriutions can become important only in the ultraclean regime where $\tau$ is very long [Fig. 2(b)].[64] A tight-binding model calculation by Tanaka et al.[16] has specified that a Pt sample is in the dirty regime when its electrical conductivity ($\sigma_{xx} \propto \tau$) is smaller than $\approx 3.3 \times 10^6$ $\Omega^{-1}$ m$^{-1}$ (or resistivity $\rho_{xx} > 30$ $\mu\Omega$ cm), which suggests that most sputter-deposited thin-film Pt would be in the dirty regime. As a result, the intrinsic spin Hall conductivity of Pt, while being almost constant in the clean limit, is calculated to decrease rapidly with decreasing $\sigma_{xx}$ or shortening $\tau$ in the dirty regime.

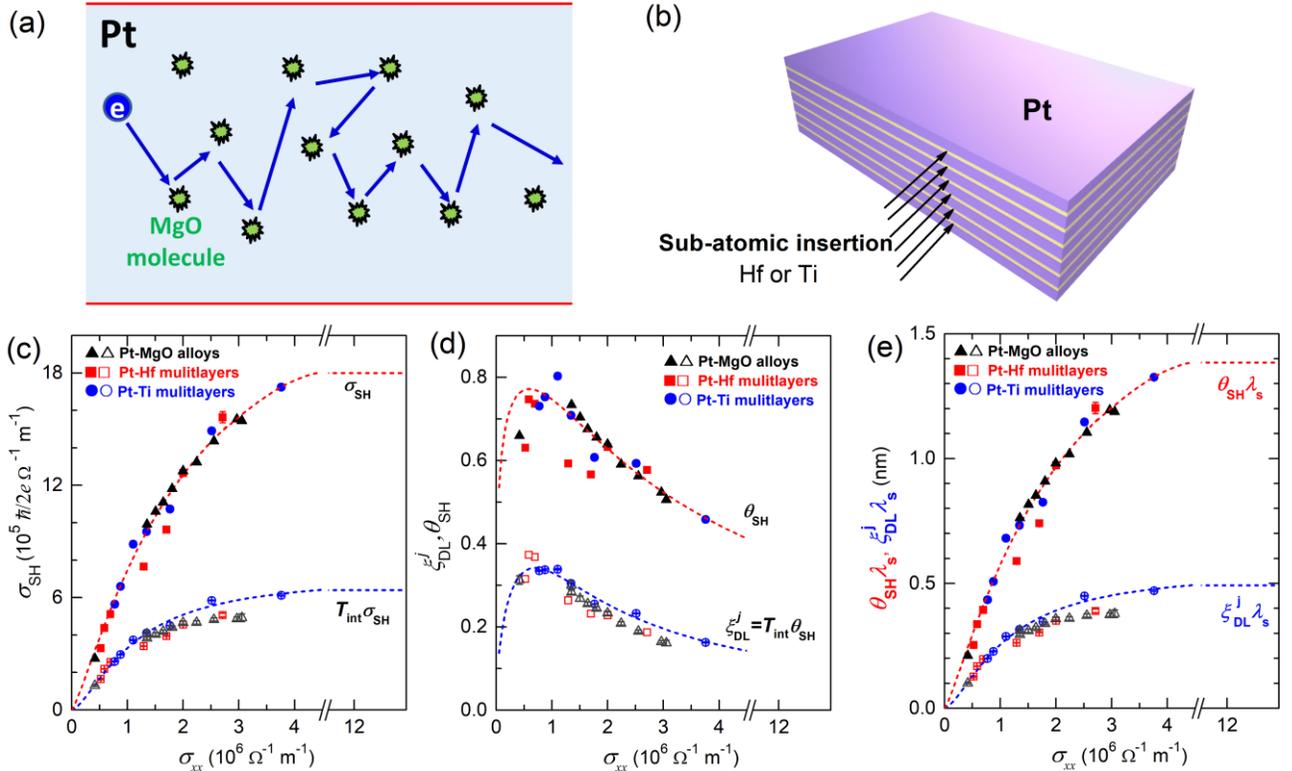

**FIG. 3.** Consequences of intrinsic spin Hall conductivity of Pt. (a) Introduction of impurity scattering by uniform MgO molecular doping. (b) Introduction of interfacial scattering by the insertion of sub-atomic spacer layers of Hf or Ti. (c) The measured ($T_{int}\sigma_{SH}$, open symbols) and internal spin Hall conductivity ($\sigma_{SH}$, solid symbols), (d) the spin Hall ratio ($\theta_{SH}$, solid symbols), damping-like SOT efficiency per unit current density ($\xi_{DL}^j$, open symbols) and (e) $\theta_{SH}\lambda_s$, $\xi_{DL}^j\lambda_s$ for the three material systems of Pt-MgO alloys, Pt-Hf multilayers, and Pt-Ti multilayers plotted as a function of the electrical conductivity ($\sigma_{xx}$). (c) is reprinted with permission from Zhu et al., Phys. Rev. Lett. 126, 107204 (2021). Copyright 2021. American Physical Society.

This expected characteristic variation of intrinsic spin Hall conductivity with carrier lifetime (or $\sigma_{xx}$) has been experimentally confirmed.[33,65,66,70] $\sigma_{xx}$ of Pt can be decreased by a factor of 8 by introducing strong impurity scattering via MgO molecular doping [Fig. 3(a)] or by introducing strong interfacial scattering via the insertion of sub-atomic spacer layers of Hf or Ti [Fig. 3(b)].[65,66] Figure 3(c) shows the measured ($T_{int}\sigma_{SH}= (\hbar/2e)\xi^E_{DL}$, open symbols) and internal spin Hall conductivity ($\sigma_{SH}=(\hbar/2e)\xi^E_{DL}/T_{int}$, solid symbols) for Pt-MgO alloys,[33] Pt/Hf multilayers,[65] and Pt/Ti multilayers[66] (for which the spin Hall conductivity is predominantly from the Pt) as a function of $\sigma_{xx}$. Here $T_{int}$ is the interfacial spin transparency estimated from Eq. (7), and $\xi^E_{DL}$ is determined from angle-dependent in-plane harmonic Hall voltage response (HHVR) measurement with any thermoelectric effects being carefully taken into account.[33,65,66] $\sigma_{SH}$ decreases rapidly towards zero with decreasing $\sigma_{xx}$ (shortening $\tau$) in the dirty regime. This observation provides direct experimental evidence that the intrinsic SHE is the dominant mechanism of the spin current generation in Pt. More detailed temperature dependent SOT measurements have further excluded any important role of the extrinsic mechanisms.[33]

### B. Spin Hall ratio and spin-orbit torque efficiency

A direct consequence of the intrinsic nature of the spin Hall conductivity of Pt is the strong, non-monotonic dependence of $\theta_{SH} = \sigma_{SH}/\sigma_{xx}$ and thus $\xi^j_{DL}$ on the electrical conductivity as shown in Fig. 3(d). When $\sigma_{xx}$ is decreased by introducing strong impurity scattering[33] or interfacial scattering,[65,66] $\theta_{SH}$ of Pt increases gradually and approaches the peak value at conductivity of $1\times10^6$ $\Omega^{-1}$ m$^{-1}$ ($\rho_{xx} \approx 100$ μΩ cm), before eventually decreasing quickly further into the dirty regime. The peak value of $\theta_{SH} \approx 0.8$ appears to represent the upper limit for Pt-based HMs[66] (where the SHE of Pt is the only important source of charge-spin conversion), which is set by the characteristic trade-off between $\sigma_{SH}$ and the carrier lifetime.[16,64] This effect explains well the enhancement of $\xi^j_{DL}$ in Pt/FM bilayers induced by doping of Al, Hf,[71] or other defects[72] into Pt. However, this is different from the cases of Au-Pt, Pd-Pt, and Cu-Pt alloys. Alloying Pt with Au,[73] Pd,[54] or Cu[74,75] can enhance both $\rho_{xx}$ and $\sigma_{SH}$ and result in maximization of $\xi^j_{DL}$ at resistivities (57.5[54] or 83 μΩ cm[73,75]) of well below 100 μΩ cm. These features suggest that the band structures of the optimized Au$_{0.25}$Pt$_{0.75}$, Pd$_{0.25}$Pt$_{0.75}$, and Cu$_{0.43}$Pt$_{0.57}$ alloys can contribute greater $\sigma_{SH}$ than pure Pt. A theory has predicted that properly alloying Pt with Au can increase the intrinsic spin Hall conductivity (Supplementary Materials of Ref. 76).

### C. Effective inverse Edelstein effect length

The intrinsic nature of the spin Hall conductivity of Pt samples also affects its "effective inverse Edelstein effect length ($\lambda_{IEE}$)", a parameter characterizing the efficiency of spin-to-charge conversion in magnetic heterostructures.[77] $\lambda_{IEE}$ is the ratio between the two-dimensional horizontal charge current density and the injected vertical three-dimensional spin current density in the charge-to-spin conversion process at Rashba or topological insulator interfaces.[77,78] When the bulk ISHE of a spin-orbit layer converts the injected spin current, e.g., by spin pumping or spin Seebeck effect, into a charge current, $\lambda_{IEE}$ can be estimated as $\theta_{SH}\lambda_s$, with $\lambda_s$ being the spin diffusion length of the spin-orbit layer. Combining the relations $\theta_{SH} = \sigma_{SH}/\sigma_{xx}$ (the SHE) and spin conductance $G_{Pt} = \sigma_{xx}/\lambda_s \approx 1.3\times10^{15}$ $\Omega^{-1}$ m$^{-2}$ (predicted to be approximately constant for the Elliot-Yafet spin-relaxation mechanism[79,80] in Pt[81]) one obtains

$$\theta_{SH}\lambda_s = \sigma_{SH}/G_{Pt}, \quad (5)$$

$$\xi^j_{DL}\lambda_s = T_{int}\sigma_{SH}/G_{Pt}. \quad (6)$$

Figure 3(e) shows $\theta_{SH}\lambda_s$ and $\xi^j_{DL}\lambda_s$ as a function of $\sigma_{xx}$ for Pt-MgO alloys, Pt-Hf multilayers, and Pt-Ti multilayers. In the clean limit, $\theta_{SH}\lambda_s = 1.4$ nm for Pt, $\xi^j_{DL}\lambda_s \leq 0.7$ nm for Pt/FM bilayers. More intriguingly, because of the characteristic variation of $\sigma_{SH}$ with carrier lifetime, both $\theta_{SH}\lambda_s$ and $\xi^j_{DL}\lambda_s$ are not constant but decrease rapidly with reducing $\sigma_{xx}$ in the dirty regime. As a result, $\theta_{SH}$ of Pt can vary inversely with $\lambda_s$ only in the clean limit, whereas it decreases with shortening $\lambda_s$ much faster than an inverse manner in the dirty regime.

## III. INTERFACIAL SPIN TRANSPARENCY

### A. Less-than-unity spin transparency

As schematically shown in Fig. 4(a), in a standard SOT process, a spin current generated in the thin-film spin current generator (SCG), e.g. a HM, diffuses to the FM layer via the mediation by electrons and drops sharply near the magnetic interface because of the degradation of $T_{int}$ by spin backflow[82-84] and spin memory loss.[85-91] For this reason, $T_{int}$ of a magnetic interface is usually smaller than unity. Using a simplifying assumption that the results of drift-diffusion analysis[82-84] are approximately independent of the weak spin memory loss, $T_{int}$ for a SOT process can be estimated as[92]

$$T_{int} = G^{\downarrow\uparrow}_{SCG/FM} / (G^{\downarrow\uparrow}_{SCG/FM} + G_{SML} + G_{SCG}/2), \quad (7)$$

when the thickness ($d$) of the spin current generator is much greater than its $\lambda_s$. Here, $G_{SML}$ an effective conductance that parameterizes spin-memory loss of the interface, $G_{SCG} = \sigma_{xx}/\lambda_s$ the spin conductance of the spin current generator. $G^{\downarrow\uparrow}_{SCG/FM}$ is the bare spin-mixing conductance of the interface, which characterizes transfer of angular momentum between the spin current in a nonmagnetic layer and the magnetization in a neighboring ferromagnetic layer.[93,94,95] Note that $G^{\downarrow\uparrow}_{SCG/FM}$ can be finite even at interfaces of insulators (e.g. ferrimagnetic Y$_3$Fe$_5$O$_{12}$ and Fe$_3$O$_4$ or antiferromagnetic NiO)[96-98] because magnons can also mediate spin transport. Another widely used concept is the effective spin-mixing conductance, which correlates to $G^{\downarrow\uparrow}_{SCG/FM}$ and $G_{SCG}$ via[99]

$$G^{\uparrow\downarrow}_{eff} = G^{\uparrow\downarrow}_{SCG/FM}/(1+2G^{\uparrow\downarrow}_{SCG/FM}/G_{SCG}). \quad (8)$$

The value of $G^{\uparrow\downarrow}_{eff}$ for a magnetic interface (e.g. HM/FM) can be, in principle, determined from the enhancement of damping by spin pumping.[44,96,97] Equation (7) can be rewritten as

$$T_{\text{int}} = T_{\text{int}}^{\text{SBF}} \times T_{\text{int}}^{\text{SML}}, \quad (9)$$

with

$$T_{\text{int}}^{\text{SBF}} = G_{\text{SCG/FM}}^{\uparrow\downarrow}/(G_{\text{SCG}}/2 + G_{\text{SCG/FM}}^{\uparrow\downarrow}) \quad (10)$$

$$T_{\text{int}}^{\text{SML}} \equiv [1 + G_{\text{SML}}^{\uparrow\downarrow}/(G_{\text{SCG}}/2 + G_{\text{SCG/FM}}^{\uparrow\downarrow})]^{-1} \quad (11)$$

where $T_{\text{int}}^{\text{SBF}}$ and $T_{\text{int}}^{\text{SML}}$ are the spin transparencies set by the spin backflow and by the spin memory loss, respectively. Equation (10) can be further rewritten as

$$T_{\text{int}}^{\text{SBF}} = 2\, G_{\text{eff}}^{\uparrow\downarrow}/G_{\text{SCG}}. \quad (12)$$

When $d$ of the spin current generator layer is not much greater than its $\lambda_s$, the spin transparency set by the spin backflow should be further modified as[81]

$$T_{\text{int}}^{\text{SBF}} = [1 - \text{sech}(d/\lambda_s)]/[1 + G_{\text{SCG}}\tanh(d/\lambda_s)/2G_{\text{SCG/FM}}^{\uparrow\downarrow}]. \quad (13)$$

Using the values $G_{\text{Pt}} \approx 1.3 \times 10^{15}\ \Omega^{-1}\ \text{m}^{-2}$ as determined in recent spin-orbit torque[81] and spin valve experiments[34,100] and $G_{\text{Pt/FM}}^{\uparrow\downarrow}$ of $0.31 \times 10^{15}\ \Omega^{-1}\ \text{m}^{-2}$ (FM = Co, $\text{Ni}_{81}\text{Fe}_{19}$, or $\text{Fe}_{60}\text{Co}_{20}\text{B}_{20}$),[101] Eq. (13) would predict $T_{\text{int}}^{\text{SBF}} < 0.5$ for metallic Pt/FM interfaces, Consequently, the optimized Pt-based elemental film,[101] alloys[33] and multilayers[65,66] that can have giant $\theta_{\text{SH}}$ of up to $\approx 0.6\text{-}0.8$ only provide $\xi_{DL}^{j}$ of <0.4 ($\xi_{DL}^{j}$ = 0.16-0.22 for pure Pt[101-103]), as indicated by the big gap between the measured and internal values in Fig. 3(d). Since drift-diffusion model has assumed an infinite spin dephasing rate in the magnetic layer,[82] caution is needed when Eqs. (8)-(13) are applied to magnetic multilayers in which the magnetic layers are thinner than its spin dephasing length (~1 nm for 3d FMs[104]). For an ultrathin FM both interfaces could affect the magnitudes of the dampinglike and fieldlike SOTs.[105,106] Therefore, a more generalized theoretical modeling of possible interplay of spin backflow, finite spin dephasing length, and spin memory loss is worthy of future efforts.

## B. Suppression of spin backflow

Spin backflow is substantial and inevitable at a diffusive Pt-based HM/FM interface when conduction electrons transport the spin angular momentum, even if $G_{\text{Pt/FM}}^{\uparrow\downarrow}$ of HM/FM interface is moderately enhanced, e.g. by the insertion of an atomic magnetic alloy layer[107,108] or by introducing interfacial spin fluctuations.[109] Very recently, insertion of a paramagnetic, insulating NiO layer has been established to effectively eliminate the spin backflow and give an effective unity spin transparency at the Pt/FeCoB interface at the optimal thickness of 0.9 nm [Fig. 4(b)].[110] In this case, the insulating NiO blocks electron-mediated spin transport and enables a thermal-magnon-assisted effective spin-mixing conductance that is close to the Sharvin conductance of Pt.[111] Consequently, the measured $\xi_{DL}^{j}$ of a Pt-Hf/NiO 0.9/FeCoB [the multilayer Pt-Hf = [Pt 0.6 nm/Hf 0.2 nm]$_5$/Pt 0.6 nm] directly reaches 0.8, the expected upper limit for the spin Hall ratio of Pt [see Fig. 5(a)].[66] Correspondingly, the directly measured $\xi_{DL}^{E}$ reaches $> 1.6 \times 10^{6}\ (\hbar/2e)\ \Omega^{-1}\ \text{m}^{-1}$ for Pt 4 nm/NiO 0.9 nm/FeCoB 1.4 nm and $0.6 \times 10^{6}\ (\hbar/2e)\ \Omega^{-1}\ \text{m}^{-1}$ for the Pt-Hf/NiO 0.9/FeCoB 1.4 nm, both of which reasonably overlap with that predicted by the drift-diffusion analysis from the samples without any NiO insertion layer.[110]

However, effective spin mixing conductance mediated by magnons is not generally greater than that mediated by electrons. While a detailed theoretical modeling is still lacking, $G_{\text{eff}}^{\uparrow\downarrow}$ of a HM/NiO/FM system should vary with the thickness of the NiO insertion layer as well as with the types of the HM and the FM because it is determined collectively by the whole "composite" HM/NiO/FM interface rather than solely by the NiO layer. This conclusion is supported by a previous spin Seebeck/inverse SHE experiment[98] that the enhancement of the spin transmission at YIG/NiO 1/HM compared to that of YIG/HM is strong when the HM is Pt but minimal when the HM is Pd or W. $G_{\text{eff}}^{\uparrow\downarrow}$ at a FM/NiO/Pt interface can be reduced to below that of the corresponding FM/Pt interface when the FM surface is oxidized during the subsequent NiO deposition and becomes a magnetically dead insulating layer that attenuates spin current.[110]

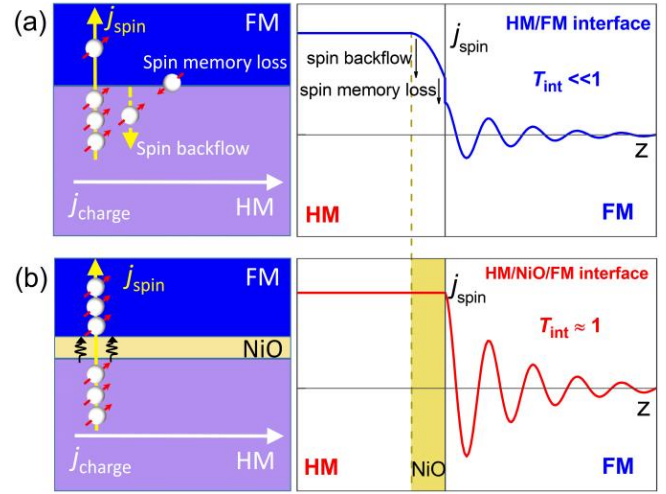

**FIG. 4** (a) Schematic of electron-mediated spin transport at a diffusive magnetic interface. The spin current diffuses from the heavy metal (HM) to the ferromagnet (FM) and undergoes substantial spin backflow and spin-memory loss ($T_{\text{int}} \ll 1$). (b) HM/NiO/FM interface where thermal magnon-mediated spin transport is free of SBF and SML ($T_{\text{int}} \approx 1$) at the optimized thickness of the paramagnetic NiO. This picture holds, at least, for the interfaces of the Pt 4 nm/NiO 0.9 nm/FeCoB 1.4 nm. Here, the spin current flows perpendicular to the layers with the transverse spins pointing perpendicular to the magnetization. In the FM the $y$ component of spin polarizations of the spin current oscillates due to the rapid precession of the spin component that is transverse to the magnetization.[82] Reprinted with permission from Zhu et al., Phys. Rev. Lett. 126, 107204 (2021). Copyright 2021. American Physical Society.

## C. Suppression of spin memory loss

The harmful effects of spin memory loss on SOTs can also be reduced by interface engineering to control the interfacial spin-orbit coupling (SOC). For a given HM/FM interface, the strength of the interfacial SOC is linearly correlated to the interfacial magnetic anisotropy energy density ($K_s^{ISOC}$) or interfacial anomalous Hall conductivity of the Pt/FM interface (see Refs. 86 and 112 for detailed discussions). When the ISOC of the same Pt/Co and Au$_{1-x}$Pt$_x$/Co interfaces is tuned by thermal engineering of spin-orbit proximity effect, $T_{int}^{SML}$ is found to decrease approximately linearly with increasing strength of interfacial SOC[110] [Fig. 5(b)], indicating the interfacial spin-orbit scattering as the dominant mechanism of spin memory loss (i.e. loss of spin memory to the lattice). In this context, interfaces such as annealed Pt/Co interfaces with strong interfacial SOC substantially degrade the SOTs. We note that, for different magnetic interfaces, spin memory loss is not always a linear function of $K_s^{ISOC}$. As predicted by the Bruno's model,[113,114] the magnitude of $K_s^{ISOC}$ of a HM/FM interface depends not only on the interfacial SOC, but also the orbital moment anisotropy and the type and thickness of the magnetic layer.[112] Meanwhile, in addition to interfacial SOC, the momentum scattering rate can also play a role in spin memory loss assuming the Elliot-Yafet relaxation mechanism[79,80] for spin-orbit scattering. The interfacial SOC and the spin memory loss can be effectively suppressed by the insertion of an ultrathin passivating layer (e.g. Hf[110,115] or Ti[116]) into the Pt/FM interfaces. Figure 5(c) shows the example of a substantial suppression of the spin memory loss at Pt/Co interface by the insertion of an ultrathin Hf layer.[110]

Interface alloying (intermixing) can also reduce the interfacial SOC and hence suppress spin memory loss.[107] An example is shown in Fig. 5(d), which indicates that intermixing at the Pt/Co interface reduces interfacial magnetic anisotropy energy density and thus the interfacial SOC and strengthens the damping-like SOT arising from the SHE in the Pt. A previous theoretical suggestion that disordered interfaces might enhance spin memory loss,[85] but in that treatment the interface alloying was *assumed* to increase rather reduce the interfacial SOC. Experiments have also clarified that magnetic proximity effect of the Pt/Co[117] and the Pt/Co$_2$FeAl interfaces[118] have negligible influence on the spin memory loss despite the controversies in the early studies[108,119-121].

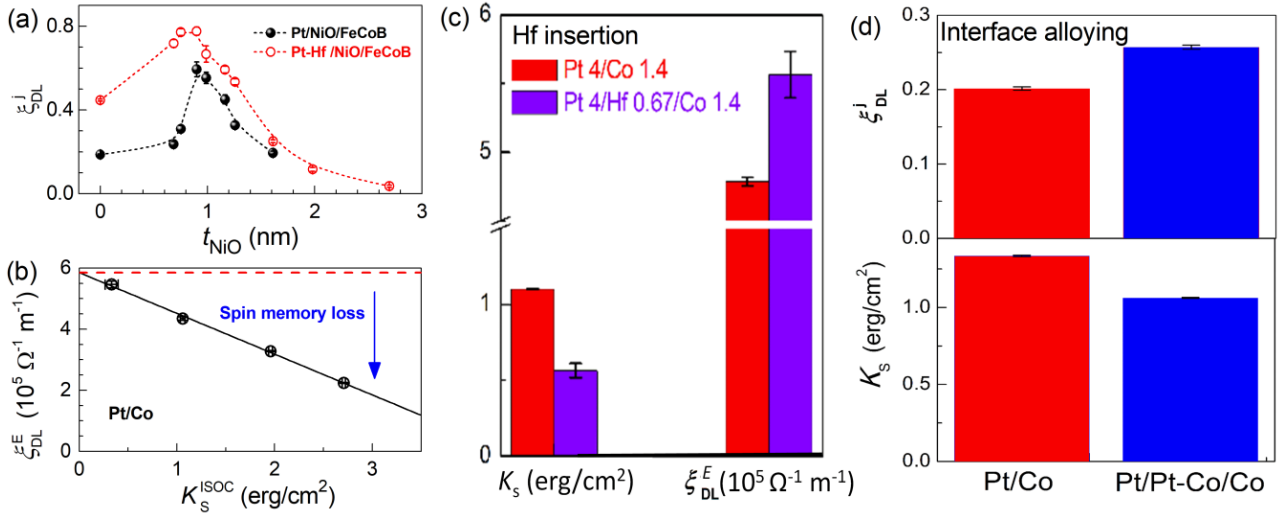

**FIG. 5.** (a) $\xi_{DL}^j$ for a Pt 4 nm/NiO $t_{NiO}$/FeCoB 1.4 nm series of samples and a Pt-Hf /NiO $t_{NiO}$/FeCoB 1.4 nm series of samples (Pt-Hf = [Pt 0.6 nm/Hf 0.2 nm]$_5$/Pt 0.6 nm) plotted as a function of the NiO thickness ($t_{NiO}$), indicating substantial suppression of spin backflow at the optimal NiO thickness of ≈ 0.9 nm. Reprinted with permission from Zhu *et al.*, Phys. Rev. Lett. 126, 107204 (2021). Copyright 2021. American Physical Society. (b) Linear reduction of $\xi_{DL}^E$ with the interfacial perpendicular magnetic anisotropy energy density ($K_s^{ISOC}$) for in-plane magnetized Pt 4 nm/Co 3.2 nm. The interfacial SOC and $K_s^{ISOC}$ are tuned by thermal engineering of the spin-orbit proximity effect at the Pt/Co interface. Reprinted with permission from Zhu *et al.*, Phys. Rev. Lett. 123, 057203 (2019). Copyright 2019. American Physical Society. (c) The total interfacial perpendicular magnetic anisotropy energy density ($K_s$) and $\xi_{DL}^E$ for Pt 4 nm/Co 1.4 nm and Pt 4 nm/Hf 0.67 nm/Co 1.4 nm, indicating a substantial reduction of $K_s$ and a significant enhancement of $\xi_{DL}^E$ by the insertion of the ultrathin Hf layer. Reprinted with permission from Zhu *et al*. Phys. Rev. Lett. 122, 077201 (2019). Copyright 2019. American Physical Society, (d) $K_s$ and $\xi_{DL}^j$ of Pt 4 nm/Co 2 nm with and without an interface alloying layer of Pt-Co = Pt$_3$Co 0.2 nm/PtCo 0.2 nm/PtCo$_3$ 0.2 nm), indicating reduction of $K_s$ and enhancement of $\xi_{DL}^j$ due to the interface alloying. Reprinted with permission from Zhu *et al.*, Phys. Rev. B 99, 180404 (2019). Copyright 2019. American Physical Society. The spin-orbit torque efficiencies in (a)-(d) are determined from harmonic Hall voltage response measurements. In (b)-(d), the CGS unit of erg/cm$^2$ is equal to the SI unit of $10^{-3}$ J/m$^2$.

**Table 1**. Representative reports of the spin Hall ratio ($\theta_{SH}$), the damping-like SOT efficiency per unit bias current density ($\xi_{DL}^j$), the damping-like SOT efficiency per applied electric field ($\xi_{DL}^E$), and the resistivity ($\rho_{xx}$) of Pt-based magnetic bilayers. ST-FMR represents for spin-torque ferromagnetic resonance, HHVR for harmonic Hall voltage response, SOT-MTJ for spin-orbit torque magnetic tunnel junction, respectively.

| Materials | $\theta_{SH}$ | $\xi_{DL}^j$ | $\xi_{DL}^E$ ($10^5$ $\Omega^{-1}$ m$^{-1}$) | $\rho_{xx}$ ($\mu\Omega$ cm) | Techniques | Ref. |
|---|---|---|---|---|---|---|
| Ni$_{81}$Fe$_{19}$/Pt 6 nm | - | 0.055 | 2.75 | 20 | ST-FMR (SOT) | Liu et al.[3] |
| Pt 4 nm/CoFe | - | 0.15 | 4.5 | 33 | ST-FMR (SOT) | Pai et al.[99] |
| Pt 4 nm/Co | - | 0.16 | 4.44 | 36 | out-of-plane HHVR (SOT) | Garello et al.[103] |
| Pt 4 nm/Co | >0.5 | 0.2 | 4.0 | 50 | out-of-plane HHVR (SOT) | Zhu et al.[86] |
| Pt 5 nm/CoTb | - | 0.22 | - | - | out-of-plane HHVR (SOT) | Wang et al.[102] |
| Pt 4 nm/Co | - | 0.15 | - | - | hysteresis loop shift (SOT) | Pai et al.[122] |
| Pt 4 nm/Fe$_{60}$Co$_{20}$B$_{20}$ | - | 0.12 | 4.8 | 25 | in-plane SOT-MTJ (SOT) | Nguyen et al.[115] |
| Ni$_{81}$Fe$_{19}$/Cu/Pt 20 nm | 0.04 | - | - | 9.4 | lateral spin valve (ISHE) | Sagasta et al.[34] |
| Ni$_{81}$Fe$_{19}$/Cu/Pt 20 nm | 0.21 | - | - | 56 | lateral spin valve (ISHE) | Sagasta et al.[34] |
| Ni$_{81}$Fe$_{19}$/Pt 12 nm | 0.12 | - | - | - | spin pumping (ISHE) | Ostbaum et al.[123] |
| Y$_3$Fe$_5$O$_{12}$/Pt 4.1 nm | 0.10 | - | - | 48 | spin pumping (ISHE) | Wang et al.[124] |
| Y$_3$Fe$_5$O$_{12}$/Pt 1-22 nm | 0.075 | - | - | - | spin Hall magnetoresistance | Althammer et al.[125] |

## IV. DISCREPANCIES OF SPIN-ORBIT TORQUE EFFICIENCY

Before discussing the discrepancies, we must first properly distinguish $\xi_{DL}^j$, $\theta_{SH}$ and so-called "effective spin Hall angle ($\theta_{SH}^{eff}$)", all of which are used in the literature. In a SOT experiment, $\xi_{DL}^j$ of a magnetic heterostructure is sometimes called "effective spin Hall angle" (i.e., $\theta_{SH}^{eff} = \xi_{DL}^j = T_{int} \theta_{SH}$), but is smaller than $\theta_{SH}$ of the spin current generator unless $T_{int}$ is unity. Since it could be conceptionally confusing, the use of "effective spin Hall angle" is avoided whenever possible in our discussions. In principle, $\theta_{SH}$ could be measured using inverse SHE (ISHE) or spin Hall magnetoresistance experiments if the spin-mixing conductance and the interfacial spin relaxation are known. There is no SOT or $\xi_{DL}^j$ in inverse SHE or spin Hall magnetoresistance experiments.

There are strong discrepancies in the literature concerning the values of $\xi_{DL}^j$ and $\theta_{SH}$ of Pt. As shown by the representative examples[53,86,99,102,103,122-125] listed in Table 1, there is, at least, fourfold disagreement on $\xi_{DL}^j$ for metallic Pt/FM bilayers. If these $\theta_{SH}$ values from SOT experiments are compared with that from ISHE[34,123,124] and spin Hall magnetoresistance[83,125] measurements, there is more than a factor of 10 disagreement.

### A. Variations of sample parameters

Variations in the reported values of $\xi_{DL}^j$ can be partly understood as due simply to the effects associated to the resistivity, the layer thicknesses, and the interfacial spin transparency of the varying samples. The effective resistivity ($\rho_{xx}$) of a pure Pt layer can vary significant depending on growth protocol (impurity scattering and grain boundary scattering) and the layer thickness (interfacial scattering). For example, $\rho_{xx}$ of a smooth pure Pt layer grown on a Ta seed layer[92] can vary from 20 $\mu\Omega$ cm to 100 $\mu\Omega$ cm when the thickness is reduced from 15 nm to 1.2 nm [see Fig. 6(a)]. $\rho_{xx}$ of a 4-5 nm thick Pt can also vary from 30 $\mu\Omega$ cm to 50 $\mu\Omega$ cm [see Table 1 and Fig. 6(d)]. If there is strong impurity scattering or interfacial scattering introduced, as is in the cases of Pt-MgO alloy,[33] Pt-Hf multilayers,[65] Pt-Ti multilayers,[66] $\rho_{xx}$ can be increased by much more, up to over 140 $\mu\Omega$ cm.[65] As a result, $\theta_{SH}$ (= $\sigma_{SH} \rho_{xx}$) and $\xi_{DL}^j$ of a Pt/FM bilayer grown on the same Ta seed layer can vary by a factor of 3 simply due to the resistivity variation [Fig. 6(e)]. Typically, $\sigma_{SH}$ and $\xi_{DL}^E$ vary less with resistivity than $\theta_{SH}$ and $\xi_{DL}^j$ [Fig. 3(c)]. More strikingly, $\rho_{xx}$ of the Pt grown on YIG surface has been reported to increase by a factor of 53, from 36 $\mu\Omega$ cm to 1900 $\mu\Omega$ cm, as the layer thickness is decreased from 22 nm to 1.1 nm,[125] from which a dramatic variation in $\theta_{SH}$ is expected.

An increase in the resistivity of Pt can additionally decrease the spin diffusion length, which should vary in proportion to $1/\rho_{xx}$ assuming that the Elliot-Yafet mechanism[79,80] dominate the spin relaxation in the Pt. As shown in Figs. 6(b) and 6(c), the thickness of the HM can strongly modulate the spin backflow and thus $\xi_{DL}^E$ of HM/FM bilayers, in good consistence with the prediction by Eq. (13). Moreover, the effective spin-mixing conductance of metallic magnetic interfaces can vary moderately with the interface details. A recent experiment indicates that interface alloying can result in a 30% increase in the effective spin-mixing conductance of Pt/Co interface.[107] First-principles calculations also predict a modest increase of effective spin mixing conductance by interfacial intermixing, e.g. 5% for Cu/Co interface[111] and 14% for Pt/NiFe interface,[108] at least in the absence of substantial ISOC. These effects, which increase the resistivity, the Pt layer thickness, and/or effective spin-mixing conductance of the interface, can thus suppress spin backflow at the interfaces, leading to increase in the spin transparency (see Eq. (13)) and the SOT efficiencies.

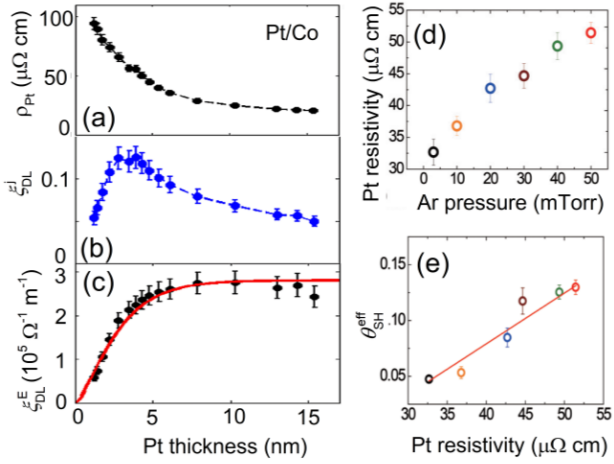

**FIG. 6**. (a) The Pt resistivity, (b) the damping-like torque efficiency per unit current density ($\xi_{DL}^{j}$), (c) the damping-like torque efficiency per electric field ($\xi_{DL}^{E}$) for a Pt 1.2-15 nm/Co 1 nm bilayer plotted as a function of the Pt layer thickness. Reprinted with permission from Nguyen *et al*., Phys. Rev. Lett. 116, 126601 (2016). Copyright 2016. American Physical Society. (d) The Pt resistivity and (c) the dependence of damping-like SOT efficiency per unit current density ($\theta_{SH}^{eff} = \xi_{DL}^{j}$) on the Pt resistivity for a Pt 5 nm/Co 0.8 nm that are sputter-deposited under different Ar pressure. Reprinted with permission from Lee *et al*., Phys. Rev. B 96, 064405 (2017). Copyright 2017. American Physical Society.

**B. Analysis of the interfacial spin transparency**

Analyses of some experiments also differ in whether they properly take into account the critical effects of scattering and relaxation of spin current at the magnetic interface. This makes a substantial difference in the obtained values of $\theta_{SH}$. A SOT experiment, in which spin current diffuses from the HM to the FM, directly measures $\xi_{DL}^{j}$. When a spin current driven by spin pumping or spin Seebeck effect in a FM is injected into, for instance, the thin-film HM, it can be detected by an ISHE voltage ($V_{ISHE}$) induced by the spin-charge conversion. Calculation of $\theta_{SH}$ from $\xi_{DL}^{j}$ in SOT experiments or from $V_{ISHE}$ in inverse SHE experiments requires accurate determination of the effective spin-mixing conductance and spin memory loss of the magnetic interface. The effective spin-mixing conductance is normally estimated based on analyzing the magnetic damping as a function of FM layer thickness under the assumption that the dominant thickness-dependent mechanism is spin pumping. However, this assumption is often incorrect for magnetic layers with in-plane anisotropy because what dominates the FM thickness ($t_{FM}$) dependence of in-plane damping ($\alpha$) of most sputter-deposited HM/FM samples in the nanometer thickness range is usually two-magnon scattering rather than spin pumping.[92,101] The signatures of the presence of a strong two-magnon scattering damping include a $1/t_{FM}^2$ scaling of in-plane damping for a wide thickness range and an unreasonably small or even negative intercept——"intrinsic damping" in a linear $\alpha$-$1/t_{FM}$ fit for the thick FM films.[101] If a strong two-magnon scattering contribution to the in-plane damping of magnetic bilayers is ignored, the effective spin-mixing conductance of the interface can be considerably overestimated,[101] resulting in underestimation of $\theta_{SH}$.

**C. Measurement techniques**

The experimental techniques used to measure SOTs can also make a big difference to the obtained $\xi_{DL}^{j}$ values. As reliable means of evaluating $\xi_{DL}^{j}$, we recommend out-of-plane HHVR measurements of strong perpendicular magnetic anisotropy (PMA) samples with square out-of-plane magnetization and Hall resistance hysteresis loops,[65,73] current-dependent shifts in out-of-plane switching field in PMA samples according to the careful procedure of ref. 122, angle-dependent HHVR measurements of in-plane anisotropy samples,[65,66,73,126] or antidamping switching measurements of in-plane magnetized 100-nm scale magnets in magnetic tunnel junstions.[8,65] While the detailed reasons still remain unclear, $\xi_{DL}^{j}$ obtained from a standard FM thickness-dependent spin-torque ferromagnetic resonance (ST-FMR) measurement[99] that has carefully taken into account field-like SOT and spin pumping is sometimes significantly smaller than that from HHVR measurement.[92]

Erroneous results of $\xi_{DL}^{j}$ can be obtained in these circumstances: (i) when an important field-like torque,[99] spin pumping,[127] and/or shape anisotropy (nanostrip only)[128] are ignored in a lineshape analysis of a ST-FMR measurement, (ii) when significant two-magnon scattering[92,101] and/or thermoelectric effects (due to radio-frequency[129] or direct currents) affect the FMR linewidth or apparent damping in a direct current biased ST-FMR measurement, (iii) when a large "planar Hall correction" is applied in an out-of-plane HHVR measurement (especially when the ratio of the planar Hall voltage to the anomalous Hall voltage is > 0.1, see the Supplementary Materials of Refs. 54, 73 and 130), (iv) when non-negligible thermoelectric effects are ignored in an in-plane[126,131] or out-of-plane[132] HHVR measurement, and (v) when the macrospin approximation failed in a HHVR measurement (the indication can be robust multi-domain features in the out-of-plane field dependence of the first HHVR of in-plane anisotropy samples or an obvious deviation of the first HHVR of a PMA sample from a parabolic scaling with in-plane magnetic fields even when the fields are small). Generally, any HM thickness-dependent analyses on SOT, ISHE, or spin Hall magnetoresistance must carefully take into account the effects accompanying the variation of resistivity with layer thickness, e.g. the thickness dependences of $\theta_{SH}$, the spin-diffusion length, and the current distribution.

In addition, when magnetic conductors (either metals[133-137] or semiconductors[138]) are used as the spin injectors in spin pumping experiments, the FMR microwave exciting magnetization dynamics also induces a number of parasitic effects simultaneously, including the dc electromotive forces due to rf-current rectification via anisotropic magnetoresistance effect, planar Hall effect, and anomalous Hall effect. Some such effects manifest with the same symmetry as that of spin-to-charge conversion, thus potentially leading to incorrect interpretations.[134,135,139,140,]

**D. Current density of magnetization switching**

SOT-driven switching of perpendicularly magnetized Hall bars in the micrometer or sub-micrometer scales cannot provide a quantitative estimation of $\xi_{DL}^j$ and $\theta_{SH}$, even though such switching experiments are of interest for advancing the understanding of magnetization reversal as well as domain wall depinning. In simplified models, $\xi_{DL}^j$ for a heterostructure with PMA is expected to correlate inversely to the critical switching current density ($j_{c0}$) in the SOT provider via Eq. (15) in the macrospin limit[141] and via Eq. (14) in the domain wall depinning regime,[142]

$$j_{c0} = e\mu_0 M_s t_{FM} (H_k - \sqrt{2}|H_x|)/\hbar\xi_{DL}^j, \quad (14)$$

$$j_{c0} = (4e/\pi\hbar) \mu_0 M_s t_{FM} H_c / \xi_{DL}^j, \quad (15)$$

where $\mu_0$ the permeability of vacuum, $H_x$ the applied field along the current direction, $M_s$, $t_{FM}$ and $H_k$ are the magnetization, thickness, and the perpendicular magnetic anisotropy field of the driven magnetic layer. However, recent experiments[143] have shown that there is no simple correlation between $\xi_{DL}^j$ and the critical switching current density of realistic perpendicularly magnetized spin-current generator/FM heterostructures. The macrospin analysis simply does not apply for the switching dynamics of micrometer-scale samples so that the values of $\xi_{DL}^j$ determined using the switching current density and Eq. (14) can produce overestimates by up to thousands of times ($\xi_{DL,macro}^j$ and $\xi_{DL,macro}^j/\xi_{DL}^j$ in Table 2).[143] A domain-wall depinning analysis [Eq. (15)] can either under- or over-estimated $\xi_{DL}^j$ by up to tens of times ($\xi_{DL,DW}^j$ and $\xi_{DL,DW}^j/\xi_{DL}^j$ in Table 2).[143] Caution is thus required to interpret data associated to perpendicular magnetization switching current density. For the same reason, so-called "switching efficiency", defined as the $H_c/j_{c0}$ ratio or the $H_k/j_{c0}$ ratio, can be highly misleading when used as a comparative guidance of $\xi_{DL}^j$ for different materials. This is particularly so when the switched FMs have significantly varying magnetic anisotropy, Dzyaloshinskii–Moriya interaction (DMI), magnetic damping, layer thicknesses, thermal stability, pinning field, saturation magnetization, etc.

In contrast, the simple macrospin model appears to reasonably approximate the antidamping torque switching of nanoscale in-plane magnetized MTJs. As compared in Table 3, the values of $\xi_{DL}^j$ estimated from the *dc* critical switching current densities of in-plane-magnetized MTJs (after ramp rate measurements, see Ref. 8 for the details of the estimation) are consistent with those from HHVR or ST-FMR techniques.

**Table 2**. Comparison of spin-torque efficiencies determined from harmonic response ($\xi_{DL}^j$) and magnetization switching ($\xi_{DL,DW}^j$, $\xi_{DL,macro}^j$) of perpendicularly magnetized bilayers. The values of $\xi_{DL,DW}^j$ are determined from a model of a current-induce effective field acting on domain walls (Eq. (15)), and $\xi_{DL,macro}^j$ is determined within a macrospin model (Eq. (14)). The $\xi_{DL}^j$ results for the Pt 6 nm/Fe$_3$GeTe$_2$ 4 nm and Ta 5 nm/Tb$_{20}$Fe$_{64}$Co$_{16}$ 1.8 nm samples were reported in refs. 144 and 145, while the corresponding values of $\xi_{DL,DW}^j$ and $\xi_{DL,macro}^j$ were calculated in ref. 143 using the reported switching current density $j_c$ values and other sample parameters as reported in ref. 144 and 145.

| Samples | $j_c$ (10$^7$ A/cm$^2$) | $\xi_{DL}^j$ | $\xi_{DL,DW}^j$ | $\xi_{DL,macro}^j$ | $\xi_{DL,DW}^j/\xi_{DL}^j$ | $\xi_{DL,macro}^j/\xi_{DL}^j$ |
|---|---|---|---|---|---|---|
| Pt 2 nm/Co 1.4 nm (annealed) | 8.2 | 0.15 | 0.48 | 2.8 | 3.2 | 18.7 |
| Pt 4 nm /Co 0.75 nm | 3.2 | 0.21 | 0.23 | 6.0 | 1.1 | 28.6 |
| [Pt 0.6 nm/Hf 0.2 nm]$_5$/Pt 0.6 nm /Co 0.63 nm | 2.4 | 0.36 | 0.38 | 3.8 | 1.1 | 10.6 |
| Pt$_{0.75}$Pd$_{0.25}$ 4 nm /Co 0.64 nm | 2.6 | 0.26 | 0.36 | 4.8 | 1.4 | 18.5 |
| Au$_{0.25}$Pt$_{0.75}$ 4 nm /Co 0.64 nm | 1.7 | 0.30 | 1.05 | 5.2 | 3.5 | 17.3 |
| Pt$_{0.7}$(MgO)$_{0.3}$ 4 nm /Co 0.68 nm | 1.5 | 0.30 | 0.09 | 16.8 | 0.3 | 56 |
| Pd 4 nm /Co 0.64 nm | 3.75 | 0.07 | 0.09 | 3.0 | 1.3 | 42.9 |
| W 4 nm /Fe$_{0.6}$Co$_{0.2}$B$_{0.2}$ 1.5 nm | 0.036 | 0.4 | 4.0 | 306 | 10 | 765 |
| Pt 6 nm /Fe$_3$GeTe$_2$ 4 nm | 1.2 | 0.12 | 0.06 | 2.0 | 0.5 | 16.7 |
| Ta 5 nm /Tb$_{20}$Fe$_{64}$Co$_{16}$ 1.8 nm | 0.04 | 0.12 | 2.1 | 234 | 18 | 1950 |

**Table 3**. Comparison of the damping-like SOT efficiency per unit bias current density ($\xi_{DL}^j$) estimated from SOT switching of in-plane magnetized MTJ devices (the current distribution is analyzed within a parallel-resistor model) with those from harmonic Hall voltage response (HHVR) or spin-torque ferromagnetic resonance (ST-FMR) measurements.

| Spin Hall channel | $\xi_{DL}^j$ | |
|---|---|---|
| | In-plane SOT-MTJ | Other techniques |
| Pt 4 nm | 0.12 [115] | 0.12 (HHVR)[81] |
| $Au_{0.25}Pt_{0.75}$ 5 nm | 0.30 [8] | 0.3-0.35 (HHVR)[73] |
| $Pt_{0.85}Hf_{0.16}$ 6 nm | 0.23 [8] | 0.23 (HHVR)[71] |
| W | -0.33 [6] | -0.3 (ST-FMR)[6] |
| Ta 6.2 nm | -0.12 [5] | -0.15 (ST-FMR)[5] |

### E. Interface-generated torques

In principle, various interface-generated SOTs can add to or subtract from the spin Hall SOT efficiencies of HM/FM samples. However, the strong discrepancies in the $\xi_{DL}^j$ values seem unlikely to be attributed to any interfacial generation of spin current or spin accumulation. A recent spin-orbit torque experiment[146] has established clean evidence that, under the same electric field, the efficiency of spin current generation by interfacial SOC effect (e.g. Rashba-Edelstein(like) effect or spin filtering effect) should be at least 2-3 orders of magnitude smaller than that of Pt even when the interfacial SOC is quite strong (e.g. at Ti/FeCoB interfaces). The absence of a significant interfacial spin current generation is well supported by: (i) a spin Seebeck/ISHE experiment that spin current generation is absent at $Bi/Ag/Y_3Fe_5O_{12}$ and $Bi/Y_3Fe_5O_{12}$ interfaces[139,140] prepared by different techniques; (ii) the universal observation of strong scaling of the SOT efficiencies or the effective spin Hall angle with the resistivity and the layer thickness of HMs,[33,34,65,66,74,81,147] topological insulators,[35] nonmagnetic complex oxides;[148] and (iii) the clear dependence of inverse spin Hall voltage on the spin-mixing conductance of the interfaces of magnetic oxides (e.g. $Y_3Fe_5O_{12}$) in spin Seebeck[44,139,140] and spin pumping.[149] This is consistent with the theories[82,150-152] that interfacial SOC has negligible contribution to the damping-like SOT via the two-dimensional Rashba-Edelstein(-like) effect of magnetic interfaces.

## V. COMPARISON OF STRONG SPIN-CURRENT GENERATORS

After a decade of intensive studies, strong spin-current generation by SOC effects has been demonstrated in heavy metals,[5-9,33,34,147] complex oxides,[148,153] topological insulators[35,18,19,154,155] and semimetals[36], and ferromagnets.[37] To compare the effectiveness of the various spin-current generators for manipulation of *nanoscale* magnets, we plot $\xi_{DL}^j$ for the representative strong spin-current generators as a function of $\rho_{xx}$ in Fig. 7(a). Notably, the high $\xi_{DL}^j$ values provided by BiSb,[35] BiSe,[155] $IrO_2$,[153] or $SrIrO_3$ [148] largely rely on their high resistivities.

In Fig. 7(b) we plot the relative values of the estimated write power and write current for SOT-MRAM devices based on the various strong spin current generators listed in detail in Table 4. In this estimation, the switching barrier of the MRAM, the spin Hall channels (length $L$ = 400 nm), the FM free layer (resistivity $\rho_{FM} \approx 130$ μΩ cm, width $W$ = 30 nm, and thickness $t$ = 2 nm, see the carton in Fig. 7(b)), and the switching current duration are assumed to be the same for different spin current generators. We compute only the write power consumed within the SOT-MRAM device and not the external circuitry, because the external power consumption may vary strongly for different applications and can be negligible (e.g. in cryogenic circuits[156,157] where the superconducting transistors and superconducting cables are used). For a SOT-MRAM device with a spin Hall channel (with thickness $d$ and resistivity $\rho_{xx}$)/FM free layer (with thickness $t$ and resistivity $\rho_{FM}$), the write current is

$$I_{write} \propto (1+s)/\xi_{DL}^j, \qquad (16)$$

where $s$ is the ratio of the shunting current in the FM layer to the current flow in the spin Hall channel. The write energy of a SOT-MRAM device includes the power dissipation of the spin Hall channel and the MTJ free layer (see the Supplementary Materials of Refs. 54 and 73 for an extended discussion) and can be approximated as[66]

$$P_{write} \propto [(1+s)/\xi_{DL}^j]^2 \rho_{xx} \qquad (17)$$

because it is dominated by current flow within the channel (fabrication of the electrodes at the top of the MTJ and at the two ends of the channel typically requires $L >> W$, see the scanning transmission electron microscopy imaging of a SOT-MRAM device in Ref. 8). The estimation of $I_{write}$ and $P_{write}$ in Table 4 has employed the parallel-resistor model, which expects $s \approx t\rho_{xx}/d\rho_{FM}$, to approximate the current shunting in SOT-MRAM devices. As shown in Table 3, antidamping torque switching analysis of in-plane SOT-MRAM based on the parallel-resistor model yields $\xi_{DL}^j$ values that agree well with those from other techniques (e.g. HHVR and ST-FMR). A finite-element analysis[158] suggested that the parallel-resistor model may overestimate the $s$ value for a very narrow in-plane MTJ ($W$ = 15 nm, $t$ = 2 nm) only if the spin Hall channel was highly resistive (e.g., $s$ for 4 nm BiSe[155] with giant resistivity of 13000 μΩ cm is estimated to be 50 by the parallel-resistor model in comparison to 11.5 by the finite-element analysis, because of which the relative value of $I_{write}$ ($P_{write}$) for the 4 nm BiSe[156] might be 4 (16) times smaller than we listed in Table 4). Current shunting effect can be also relatively weaker at the corners of the MTJ free layer on a very resistive channel (the carton in Fig. 7(b)), which is suggested by the finite-element analysis[158] to lower by a factor of 2.5 the spatially averaged value of $s$ for a very resistive BiSe device ($\rho_{xx}$ = 13000 μΩ) with a small width-height ratio of $W/t$ = 7.5 (width $W$ of only 15 nm and thickness $t$ of 2 nm for the MTJ). However, that correction is expected to a minor effect for a realistic *stable* SOT-MRAM device which is necessary to have a width $W$ of > 30 nm (the reported $W$ values are 30-110 nm for in-plane Cornell devices,[8,9,32,157] 40-80 nm for in-plane Toshiba devices,[159] 120 nm for in-plane TDK devices,[160] 160 nm for in-plane Tohoku devices,[161] 60-80 nm for perpendicular IMEC devices,[162,163] and 275 nm for perpendicular CNRS devices[164]) and a small thickness $t$ (1.4-1.8 nm[8,9,32,157,159-161] for in-plane devices and ≈1 nm for perpendicular devices[7,162-164]).

It is also worth pointing out that use of the oversimplified relation

$$P_{\text{write}} \propto (1/\xi_{\text{DL}}^j)^2 \rho_{xx} \quad (18)$$

that assumes zero current shunting into the magnetic layer (i.e. $s = 0$) typically leads to erroneous estimation of the power consumption, particularly so for highly resistive spin current generators, e.g. topological insulators. We also note that the switching current densities of micrometer-sized Hall bar devices consisting of varying spin current generators and varying magnetic layers cannot provide a benchmarking for the efficiencies of the spin current generators themselves or their power performance in MRAM technologies. This is not only because the switching mechanism is distinct for large Hall bar devices (domain wall depinning and propagation) and a few tens of nm MTJs (approximately coherent rotation), but also because the switched magnetic layers usually differ considerably in terms of thickness, magnetic anisotropy, domain wall pinning, magnetization, DMI, damping, and thermal responses to the large-amplitude and long-duration switching currents.

As shown in Fig. 7(b) and Table 4, compared to the sputter-deposited clean-limit Pt ($\xi_{\text{DL}}^j = 0.055$, $\rho_{xx} = 20$ μΩ cm),[3] the optimized alloys [i.e. Au$_{0.25}$Pt$_{0.75}$,[73] Pd$_{0.25}$Pt$_{0.75}$,[54] and Pt$_{0.6}$(MgO)$_{0.4}$[33]] and multilayers (Pt-Hf = [Pt 0.6 nm/Hf 0.2 nm]$_5$/Pt 0.6 nm, Pt-Ti =[Pt 0.75 nm/Ti 0.75 nm]$_7$/Pt 0.75 nm) [65,66] are 10 times more energy efficient, or even better. The Pt/NiO and Pt-Hf multilayer/NiO[110] with improved interfacial spin transparency and reduced current shunting give a >100 times and > 50 times improvement relative to clean-limit Pt, respectively. None of the very high-resistivity sputter-deposited materials, e.g. β-Ta,[5] β-W,[6] BiSe,[155] and BiSb[35] give a benefit compared to the clean-limit Pt due to the large write resistance and the significant current shunting (if the finite-element analysis result[158] of $s = 11.5$ was used, the 4 nm BiSe device would be a factor of 2 more energy-efficient than the pure Pt device). The epitaxial Bi$_2$Se$_3$,[18] IrO$_2$,[153] and SrIrO$_3$ [148] grown by molecular beam epitaxy (MBE) or pulsed laser deposition (PLD) on single-crystalline substrates are only slightly more energy efficient than a sputter-deposited clean-limit Pt.

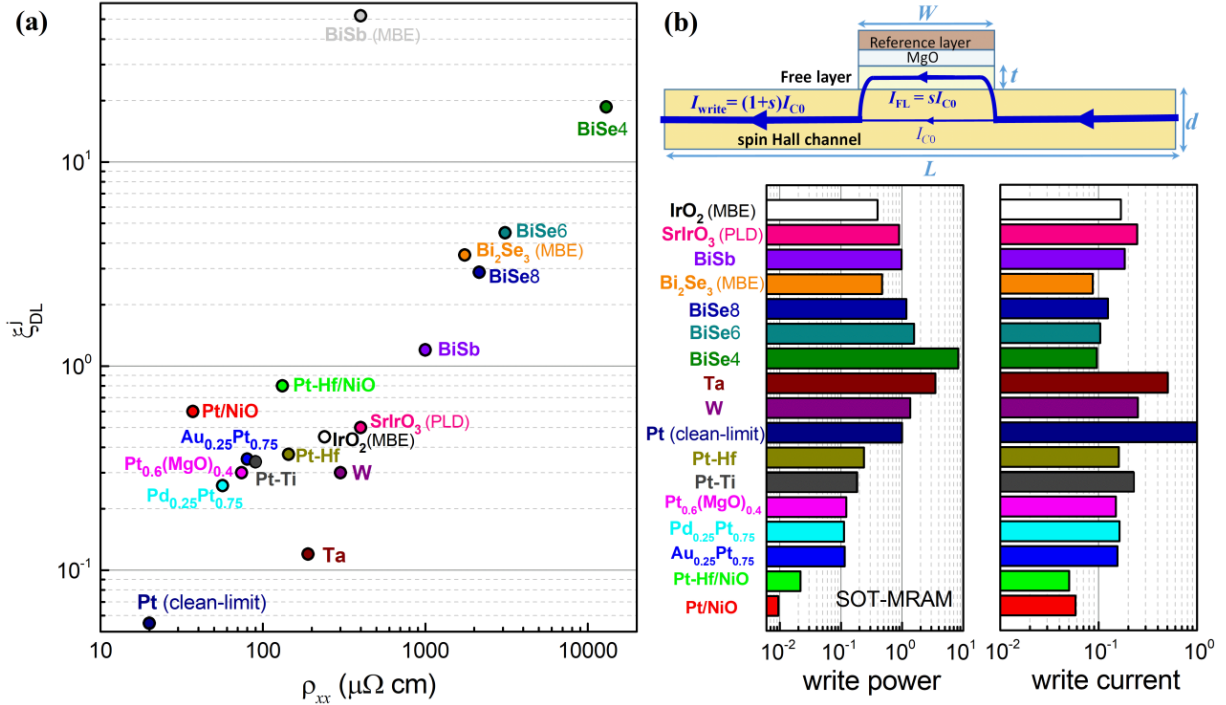

**FIG. 7**. (a) $\xi_{\text{DL}}^j$ vs $\rho_{xx}$ for representative strong spin current generators, including sputter-deposited clean-limit Pt[3], β-Ta[5], β-W[6], Pd$_{0.25}$Pt$_{0.75}$,[54] Pt$_{0.6}$(MgO)$_{0.4}$,[33] Au$_{0.25}$Pt$_{0.75}$,[73] Pt-Hf multilayers,[65] Pt-Ti multilayers,[66] Pt/NiO,[110] Pt-Hf/NiO,[110] BiSb,[35] and BiSe (4 nm, 6 nm, and 8 nm)[155], MBE-grown IrO$_2$,[153] BiSb,[154] and Bi$_2$Se$_3$,[18] and PLD-grown SrIrO$_3$.[148] (b) Cartoon of current flow in a spin Hall channel and a MTJ free layer and the estimated write power and write current ($I_{\text{write}}$) required for switching a SOT-MRAM device based on these spin current generators (plotted relative to the value of clean-limit Pt with low resistivity). $I_{c0}$ is the critical switching current in the spin Hall channel (length $L$, thickness $d$, resistivity $\rho_{xx}$) for SOT generation, while $I_{\text{FM}} = sI_{c0}$ represents the current shunting into the FM layer (width $W$, thickness $t$, resistivity $\rho_{\text{FM}}$). In (a) and (b), Pt-Hf = [Pt 0.6 nm/Hf 0.2 nm]$_5$/Pt 0.6 nm, Pt-Ti =[Pt 0.75 nm/Ti 0.75 nm]$_7$/Pt 0.75 nm. See Table 4 for more details.

**Table 4**. Comparison of the representative spin current generators. The current shunting ratio (*s*), write power, write current, and channel impedance (*R*) of their SOT-MRAM devices are estimated by assuming a parallel-resistor model. Pt-Hf = [Pt 0.6 nm/Hf 0.2 nm]$_5$/Pt 0.6 nm, Pt-Ti =[Pt 0.75 nm/Ti 0.75 nm]$_7$/Pt 0.75 nm. *s* = 0 was used for the Pt 4 nm/NiO 0.9 nm and Pt-Hf /NiO 0.9 nm since the insulating NiO should substantially suppress or even effectively block the current shunting into the thin resistive FeCoB free layer. *R* is estimated for a 400 nm long and 200 nm wide spin Hall channel.

| Material | $d$ (nm) | $\xi_{DL}^j$ | $\rho_{xx}$ (μΩ cm) | $s$ | Power | Current | $R$ (kΩ) | Ref. |
|---|---|---|---|---|---|---|---|---|
| Pt 4 nm/NiO 0.9 nm | 4 | 0.6 | 37 | 0 | 0.01 | 0.06 | 0.19 | Zhu *et al.* [110] |
| Pt-Hf /NiO 0.9 nm | 4.6 | 0.8 | 132 | 0 | 0.02 | 0.05 | 0.57 | Zhu *et al.* [110] |
| Au$_{0.25}$Pt$_{0.75}$ | 5 | 0.35 | 80 | 0.25 | 0.11 | 0.16 | 0.32 | Zhu *et al.* [73] |
| Pt$_{0.25}$Pd$_{0.75}$ | 4 | 0.26 | 56.5 | 0.22 | 0.11 | 0.16 | 0.28 | Zhu *et al.* [54] |
| Pt$_{0.6}$(MgO)$_{0.4}$ | 4 | 0.3 | 74 | 0.28 | 0.12 | 0.15 | 0.37 | Zhu *et al.* [33] |
| Pt-Ti | 7.4 | 0.34 | 90 | 0.19 | 0.18 | 0.23 | 0.24 | Zhu *et al.* [66] |
| Pt-Hf | 4.6 | 0.37 | 144 | 0.48 | 0.24 | 0.16 | 0.63 | Zhu *et al.* [65] |
| Pt (clean-limit) | 6 | 0.055 | 20 | 0.05 | 1 | 1 | 0.07 | Liu *et al.* [3] |
| β-W | 4 | 0.3 | 300 | 1.15 | 1.4 | 0.25 | 1.5 | Pai *et al.* [6] |
| β-Ta | 4 | 0.12 | 190 | 0.73 | 3.5 | 0.50 | 0.95 | Liu *et al.* [5] |
| BiSe | 4 | 18.6 | 13000 | 50 | 8.3 | 0.10 | 65 | DC *et al.* [155] |
| BiSe | 6 | 4.5 | 3100 | 7.95 | 1.6 | 0.10 | 10 | DC *et al.* [155] |
| BiSe | 8 | 2.88 | 2150 | 4.13 | 1.2 | 0.12 | 5.4 | DC *et al.* [155] |
| Bi$_2$Se$_3$ (MBE) | 8 | 3.5 | 1750 | 3.37 | 0.47 | 0.09 | 4.4 | Mellnik *et al.* [18] |
| BiSb | 10 | 1.2 | 1000 | 1.54 | 0.98 | 0.18 | 2 | Chi *et al.* [35] |
| BiSb (MBE) | 10 | 52 | 400 | 0.62 | 8.6×10$^{-5}$ | 0.003 | 0.8 | Khang *et al.* [154] |
| SrIrO$_3$ (PLD) | 8 | 0.5 | 400 | 0.77 | 0.89 | 0.25 | 1 | Nan *et al.* [148] |
| IrO$_2$ (MBE) | 5 | 0.45 | 240 | 0.74 | 0.40 | 0.17 | 0.96 | Bose et al. [153] |

For some applications in which the write impedance is predominantly from the external transistor circuits rather than from the spin Hall channel, the current efficiency ($I_{write}$) of the SOT-MRAM devices would be more critical for the total write power ($\propto I_{write}^2$) than the power consumption within the SOT-MRAM devices, because the latter could be only negligible compared to the external power consumption. As shown in Fig. 7(b) and Table 4, compared to the clean-limit Pt,[3] the Pt/NiO and Pt-Hf multilayer/NiO[110] require a factor of ~20 lower write current, while Au$_{0.25}$Pt$_{0.75}$,[73] Pd$_{0.25}$Pt$_{0.75}$,[54] Pt$_{0.6}$(MgO)$_{0.4}$,[33] Pt-Hf multilayers,[65] Pt-Ti multilayers,[66] BiSb,[154] and IrO$_2$ (MBE)[153] also require a factor of > 5 lower write current. In this case, the very high-resistivity BiSe[155] and Bi$_2$Se$_3$ (MBE)[18] are also 10 times more current-efficient than the clean-limit Pt.

In addition to energy and current efficiencies, a relatively low resistivity and integration-friendly properties of spin-current generators are also required for practical applications. A relatively low resistivity is critical for device endurance. For example, use of a large-$\rho_{xx}$ spin Hall material (e.g., 200 - 300 μΩ cm for W,[9,6] see Fig. 7(a)) will limit the endurance of SOT devices via Joule-heating-induced bursting and migration of the write line[161] as well as result in a high write impedance that is challenging to accommodate for superconducting circuits in a cryogenic computing system.[156]

The Pt-based alloys and multilayers are advantageous for their relatively low resistivities. As estimated in Table 4, SOT-MRAM devices based on the Pt-alloys can have write impedance of < 300 Ω even if a 400 nm long channel is used, which is highly desirable for the cryogenic memory applications where the superconducting electronics cannot afford a large impedance.[156,157] In contrast, the giant resistivity of BiSe or Bi$_2$Se$_3$ is expected to question the endurance of the spin Hall channels of the SOT-MRAMs. Pt-based spin current generators are also advantageous for their integration-friendly properties including the thermal and chemical stabilities, compatibility with standard sputtering deposition on SiO$_2$ substrate, and ease of being combined with standard high-performance FeCoB MTJs.[32] While the BiSb single-crystal film[154] ($d$ =10 nm, $\xi_{DL}^j$=52, $\rho_{xx}$= 400 μΩ cm) grown by MBE on single-crystal substrates does suggest a very low power/current consumption (Table 4), its usefulness in practical spin orbitronics is questioned by the low melting point of 275 ºC for BiSb.[165] This is because the integration with MTJs and CMOS circuits requires processes at much higher temperatures (typically 350 ºC). The thermal stability of thin-film BiSe is also yet to verify since some studies[166] have suggested that Bi$_2$Se$_3$ nano-crystals sublimates at temperatures of < 280 ºC.

## VI. SPIN-ORBIT-TORQUE MRAM DEVICES

One of main motivation for SOT research is its potential for enabling fast and energy-efficient non-volatile magnetic memories. Many key electronic technologies, e.g., large-scale computing, machine learning, and superconducting electronics, would benefit from the development of new fast, non-volatile, and energy-efficient magnetic memories. While the conventional non-volatile 2-terminal spin-transfer-torque magnetoresistive random access memory (STT-MRAM) with large perpendicular magnetic anisotropy [Fig. 8(a)] is attractive for its good scalability and high thermal stability[167,168] during fast sub-ns write,[169] the required high write current density limits the scaling and can exert severe stress on the MTJ and induce wear-out and breakdown of the MTJ barrier.[160] Meanwhile, the shared read/write path makes it challenging to eliminate write-upon-read errors. As an alternative, 3-terminal SOT-MRAM[5,7-9] has the potential to mitigate these issues. In a SOT-MRAM [Figs. 8(b) and 7(c)], the spin current generated by the SHE of a heavy metal layer switches the magnetic free layer of a MTJ so that the read and write paths are separated, and writing requires only low voltages. The SOT-MRAMs can have long data retention, zero standby power, and fast reliable write.[7-9,162] Experiments[8,32,170,171] and micromagnetic simulations[32,171] have established that collinear in-plane SOT-MRAM devices [Fig. 9(a)] are free of incubation delays and can be operated much faster than that predicted by some macrospin simulations.[161] While a comprehensive understanding is still lacking, the fast switching speed of the collinear in-plane SOT-MRAM devices has been suggested to be related to the accelerated micromagnetics by the pulsed Oersted field from the spin Hall channel[32,170] and the spatial magnetic non-uniformity induced by the DMI[172] and the tapering of the MTJ free layers.[8] It is an important point, however, that incubation delays showing up in some in-plane STT-MRAM devices were not generally observed in the fast anti-damping STT switching regime of in-plane all-metal spin valves,[173-175] clearly demonstrating that the pre-switching delay is neither an inherent feature of antidamping switching, nor is it purely related to thermal activation.

Figure 9(b) shows an example of fast pulse switching phase diagram (1000 events per point) of an in-plane SOT-MRAM device based on $Au_{0.25}Pt_{0.75}$ ($\rho_{xx}$ = 80 μΩ cm),[8] which indicates reliable switching at pulse widths down to 0.2 ns. Figure 9(c) shows the write error rate (WER) statistics during up to $10^5$ switching attempts per point. WERs for the pulse duration of 1 ns scale down quickly as the write voltage (current) increases, extrapolation of which indicates WERs of < $10^{-5}$ at 4 mA ($2.2 \times 10^8$ A/cm$^2$) for both the P→AP and the AP→P switching. Optimization of the magnetic damping of the device is expected to substantially reduce the current requirement in the fast pulse write process even further. Reliable sub-ns switching has also been demonstrated in in-plane SOT-MRAM devices using Pt or $Pt_{85}Hf_{15}$ as the spin Hall channel.[32,171] Here, the duration of the switching current pulses is the same or longer than the time at which the sign of the easy-axis component of the magnetization reverses due to applied current-induced antidamping torque, but it might be slightly different from the total switching time.

Collinear in-plane SOT-MRAMs can be switched directly by spin current from the spin Hall channel free of external field[170,176] as soon as the dipole field of the reference layer is diminished, e.g., by using proper synthetic anti-ferromagnetic (SAF). In-plane SOT-MRAMs also have the full freedom to adopt any beneficial SOT provider. For example, use of the low-resistivity Pt-based alloys and multilayers[33,54,65,66,73,110] can be helpful in decreasing write currents ($I_{write}$) and energies, achieving unlimited endurance (electro-immigration $\propto I_{write}^2$),[160] and also for matching the impedance of superconducting circuits in cryogenic computation systems.[156,157] Figure 9(d) compares the measured critical switching current density of representative SOT-MRAM devices based on different strong spin Hall materials. The in-plane SOT-MRAM devices based on the Pt-Hf multilayers ($\rho_{xx}$ = 130 μΩ cm)[65] show record-low critical current density of $3.6 \times 10^6$ A/cm$^2$ (the total switching current was only 73 μA for the 300 nm wide spin Hall channel) for *thermal assist-free* switching as determined by ramp rate measurements, which is significantly lower than others including the Ta- or W-based perpendicular SOT-MRAMs.[163,164]

It is true that lower switching current densities compared to the ones we have discussed for SOT-MRAM devices are possible for micrometer-sized Hall bar devices, especially in the presence of the large long-duration current-induced thermal assist [e.g. $3.6 \times 10^5$ A/cm$^2$ for perpendicular $W/Fe_{60}Co_{20}B_{20}$ ($\xi_{DL}^j$ = 0.4),[143] $4 \times 10^5$ A/cm$^2$ for perpendicular $Ta/Tb_{20}Fe_{64}Co_{16}$ ($\xi_{DL}^j$ = 0.12),[145] $2.8 \times 10^5$ A/cm$^2$ for perpendicular $Bi_2Se_3/CoTb$ ($\xi_{DL}^j$ = 0.16),[177] and $6 \times 10^5$ A/cm$^2$ for in-plane $Bi_2Se_3/Ni_{81}Fe_{19}$ ($\xi_{DL}^j$ = 1.0)[178]]. However, the performance of devices in the regime of domain-wall-mediated switching are largely irrelevant to practical technologies, because the switching currents grow by orders of magnitude when the size of magnetic layers is reduced to the sub-50-nm scale where domain wall nucleation is no longer possible.[179] In addition, the current-induced heating does not significantly assist fast operation in the ns or sub-ns scales, while nearly compensated ferrimagnets and low-anisotropy $Ni_{81}Fe_{19}$ unlikely allow for fast reliable reading using high-TMR MTJs.

One challenge in adapting in-plane SOT-MRAM devices for practical technologies is that their relatively low coercivity (< 400 Oe) and shape anisotropy limit the stabilities against external magnetic noises and thermal effects in the nanometer scale (e.g. < 50 nm). However, recent advances in the use of strain and voltage gating architecture proposed in the industry[159,180,181] have provide possible routes to enhance the coercivity and thermal stability and to reduce the transistor numbers, which suggests that in-plane SOT-MRAM technologies can be also very stable and dense. Note that the density of SOT-MRAM technologies is determined by the numbers and dimensions of the transistors used to source the switching current rather than by the MTJ nano-pillars, while the required transistor dimension is proportional to $I_{c0}$.[182] We expect that fast, efficient, reliable in-plane SOT-MRAMs are promising for both applications where high density is not a primary driver (e.g. cache memory for superconducting computation) and more general applications where reasonably high density is required (nonvolatile substitution for SRAM).

In addition to in-plane SOT-MRAM devices, high-performance perpendicular SOT-MRAMs are also of interest since a strong perpendicular magnetic anisotropy of the magnetic free layer can benefit the thermal stability in the tens of nanometer scale. However, perpendicular SOT-MRAMs face several important challenges that make it difficult for their performance to be competitive with in-plane devices. First, a highly resistive Ta or W channel is typically required to achieve perpendicular magnetic anisotropy of the magnetic free layer,[163,183] which prevents perpendicular SOT-MRAMs from benefiting from other spin-current generation materials (including the optimized Pt-alloys and multilayers[33,35,65,66,73,110]). Perpendicular SOT-MTJs also require a markedly high write current in the ns and sub-ns pulse regime[164] because the orientation of the spin polarization of the SOT from heavy metals is in the plane rather than parallel to the out-of-plane anisotropy axis. Furthermore, they require the assist of a strong in-plane effective magnetic field along the write current direction to overcome the DMI at the interface of the spin Hall channel and the magnetic free layer (interpreted in some literature as "to break the reversal symmetry" of a macrospin). So far, the required in-plane bias magnetic field has been shown to be achievable by stray field or interlayer coupling from an adjacent ferromagnetic layer,[163,184,185] exchange bias field from an adjacent antiferromagnetic layer,[186] built-in magnetic field from a lateral structural asymmetry,[187] ferroelectric field from a PMN-PT substrate,[188] or a spin transfer torque from additional large write current in the MTJ nanopillar.[183] However, the effectiveness of these strategies require experimental verification at the device level, because they may lower the scalability, the current efficiency, the reliability, or/and the endurance of the MTJ cells. Alternatively, these issues may be addressed if a thin-film system can be developed in the future to simultaneously promote strong perpendicular magnetic anisotropy, integrate with high-performance MTJs, and allow very efficient generation of a perpendicularly polarized spin current (possible only for sufficiently low-symmetry crystalline materials). However, this has remained an unmet challenge despite of the recent efforts.[189-192]

Fukami et al.[161] has also proposed orthogonal in-plane SOT-MRAM devices (so-called "x-type") with the idea that the spin current polarized orthogonally to the magnetic easy axis of the MTJ free layer might benefit short-pulse switching at the cost of the requirement of a large out-of-plane external magnetic field ($H_z$). As verified by a recent experimental study,[176] when an assist field of ~1 kOe or greater is applied, such orthogonal in-plane SOT-MRAM devices may be switched by 1 ns current pulses (and perhaps also shorter ones) at switching current densities slightly lower than that of the collinear in-plane SOT-MRAM devices (under zero applied field, Fig. 8(c)). However, for pulse durations of a few ns or longer, the orthogonal in-plane SOT-MRAM devices, even if an assist field of up to ~1.4 kOe is applied, still require a switching current density higher than that of the collinear devices.

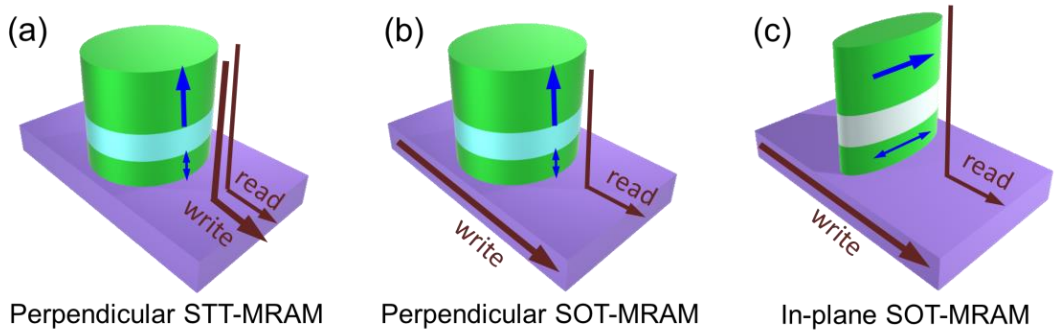

**FIG. 8**. Illustration of (a) perpendicular STT-MRAM, (b) perpendicular SOT-MRAM, and (c) in-plane SOT-MRAM.

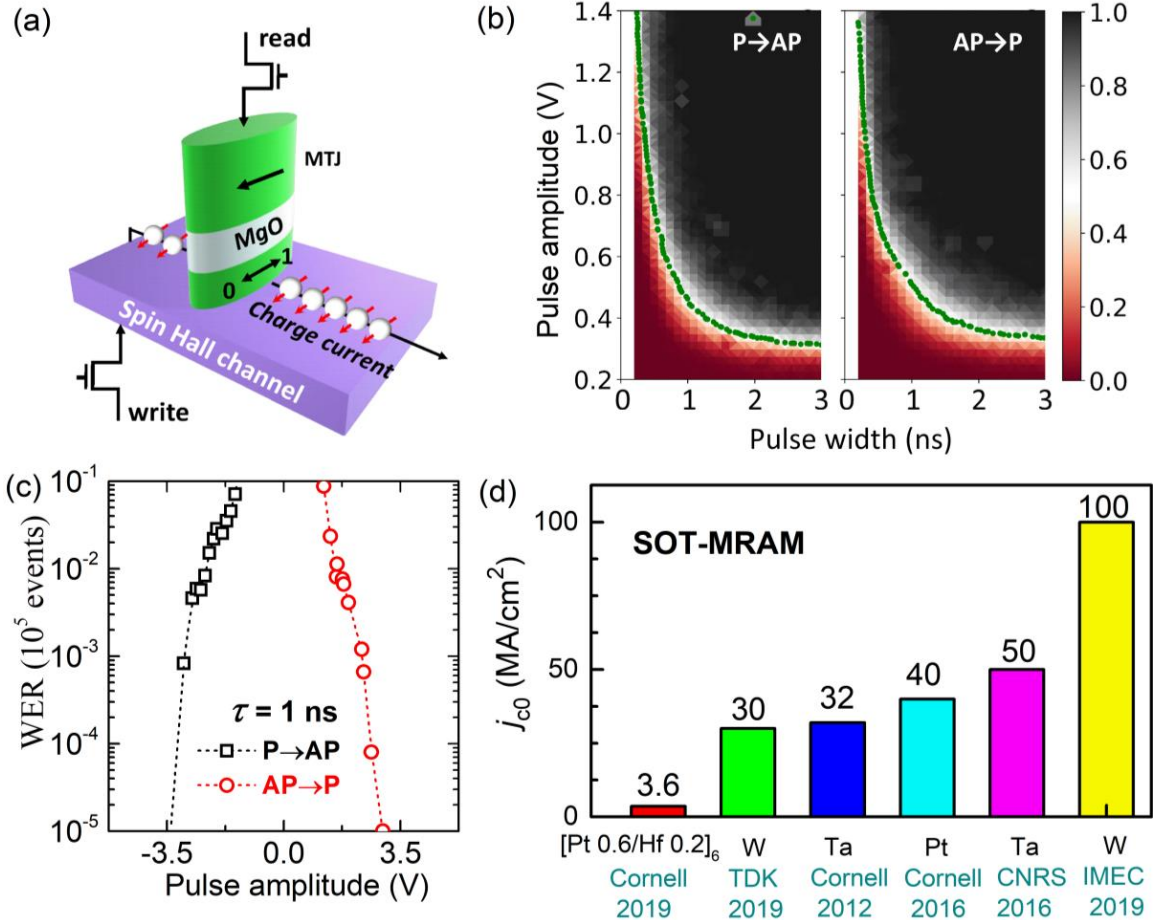

**FIG. 9**. (a) Schematic of an in-plane SOT-MRAM cell. (b) Fast-pulse switching phase diagrams of a $Au_{0.25}Pt_{0.75}$-based MRAM device (the color scale represents the switching probability for 1000 events per point), the green dots indicate the 50% switching probability points. (c) The write error rates (WERs) with 1 ns pulse ($10^5$ events per point) plotted as a function of write current for the $Au_{0.25}Pt_{0.75}$-based MRAM device. Reprinted with permission from Zhu *et al*., Adv. Electron. Mater. 6, 1901131 (2020). Copyright 2020. WILEY-VCH Verlag GmbH & Co. KGaA, Weinheim. (d) Comparison of the reported critical write current densities of SOT-MRAMs based on spin channel materials of [Pt 0.6 nm/Hf 0.2 nm]$_6$,[65] W,[160,164] Ta,[5,164] and Pt.[32]

## VII. CONCLUSION AND PERSPECTIVES

We have reviewed recent progress towards maximizing the efficiencies of charge-spin conversion, interfacial spin transmission, and thus SOT in Pt-based heavy metal/ferromagnet systems and SOT-MRAM prototypes. Compared to the wide variety of materials being considered to achieve efficient spin current generation, the newly developed Pt-based heavy metals are outstanding as energy-efficient, high-endurance, integration-friendly options. They simultaneously combines several key advantages: high damping-like SOT efficiency, the relatively low resistivity, high thermal and chemical stabilities, compatibility with standard sputtering deposition on $SiO_2$ substrate, and ease of being integrated with standard high-performance FeCoB MTJs.[32] Utilizing the $Au_{0.25}Pt_{0.75}$ alloy[8] and the Pt/Hf multilayers[65] as spin Hall channel materials, we have demonstrated reliable three-terminal SOT-MRAM devices with write speed of < 200 ps and critical switching current density of $3.6 \times 10^6$ A/cm$^2$.

Despite of these inspiring progresses, a number of essential questions remain open.

(i) Achieving accurate quantitative theoretical calculations of the spin Hall conductivity remains challenging both for Pt and for most other spin-current generating materials. As summarized in Table 5, the predicted clean-limit values of the intrinsic $\sigma_{SH}$ of Pt from available theories[16,17,76,193,194] differ by more than a factor of 10, i.e. $\sigma_{SH} = (0.48\text{-}4.56) \times 10^5$ ($\hbar/2e$) $\Omega^{-1}$ m$^{-1}$ in the clean limit ($\sigma_{xx} > 1.6 \times 10^6$ $\Omega^{-1}$ m$^{-1}$). More importantly, these theoretical values are, at least, a factor of 4 smaller than the experimental value of $>1.6 \times 10^6$ ($\hbar/2e$) $\Omega^{-1}$ m$^{-1}$ that was measured directly from the Pt 4 nm/NiO 0.9 nm/FeCoB 1.4 nm[110] without any correction of spin transparency following the drift-diffusion model. This disagreement implies that there are fundamental gaps in our understanding of the microscopic mechanisms that allow Pt-based devices to achieve efficient spin-orbit torques. To put it bluntly, substantial theoretical advances are likely still required to understand the giant SHE of Pt. A more generalized drift-diffusion model is also required to include interfacial spin memory loss and the effects of finite spin relaxation rates and layer thickness of the magnetic layers.

(ii) In addition to the bulk SHE that has been established as the dominant source of the spin current generation in Pt-based heavy metal/ferromagnet systems, there is also the possibility that other novel phenomena might be harnessed to enhance the SOT. Possibilities that have been identified theoretically and/or experimentally include spin swapping,[195-197] the spin-orbit filtering effect,[20] the orbital Hall effect,[198] "self-induced" bulk or surface SOTs,[37,63,199] impurity-driven interfacial skew scattering.[200,201] There is essential work to be done to understand whether any of these effects can actually be manipulated to generate torques comparable to or stronger than the bulk SHE, and if so, how they might best be employed together with the bulk SHE. Write currents for SOT-MRAM devices might also be further reduced by combining SOT with other manipulation strategies, such as voltage controlled magnetic anisotropy (VCMA)[181] or conventional spin transfer torque.

(iii) Pt/ferromagnet interfaces typically have strong interfacial DMI,[54,112] which raises a question as to how the performance of SOT-MRAMs is impacted by the DMI. For in-plane SOT-MRAMs, the DMI can increase micromagnetic non-uniformity,[172] leading to improvements of the switching speed,[170] and increases in magnetic damping,[101] at the cost of lower the thermal stability. Perpendicular MTJ experiments[202,203] and modeling[204-207] have shown that the rigid macrospin reversal is never accurate until the MTJ device size is quite small (< 20-50 nm). As a result, larger perpendicular MTJs are switched through thermally assisted nucleation of reversal magnetic domain and SOT-driven depinning and propagation of magnetic domain walls. In this case, interfacial DMI and magnetic damping are directly involved in the determination of the speed, the reliability, and the power consumption of SOT switching of perpendicular MTJs. Quantitative investigation of the roles of interfacial DMI and magnetic damping during the SOT-driven magnetization switching, particularly in the ns and sub-ns timescales, will provide insight onto the mechanisms that are essential for the optimization of SOT devices.

The progress we have described to improve SOT materials and devices suggests a promising future of SOT-based applications. Utilizing the strong SOT, relatively low resistivity, and the strong, tunable interfacial DMI[54,112] of the most recently developed Pt-based spin Hall metals, prototype SOT-MRAM devices have already demonstrated ultrafast, ultralow-power, and high-endurance writing. Considerable further improvement is possible by optimizing damping, DMI, and other parameters. In addition to MRAM, the new developed SOT materials and strategies might also enable terahertz emitters,[208,209] nano-oscillators,[48,210,211] and logic devices[11,212,213] that have significantly improved performance compared to the existing devices based on the conventional spin current generators like pure Pt, Ta, or W. Furthermore, the implementation of the strong SOT materials into spintronic devices can enable the development of novel beyond-CMOS functionalities, e.g. artificial synapses and neurons[214] based on superparamagnetic tunnel junctions,[215-218] nano-oscillators,[219] magnetic domain walls, and skyrmions[220] for new low-power computer architectures.

**Table 5**. Theoretical predictions of the intrinsic spin Hall conductivity ($\sigma_{SH}$) and the electrical conductivity ($\sigma_{xx}$) of Pt.

| Method | $\sigma_{SH}$ [($\hbar/2e$)$\Omega^{-1}$ m$^{-1}$] | $\sigma_{xx}$ ($\Omega^{-1}$ m$^{-1}$) | Ref. |
|---|---|---|---|
| tight-binding | $0.1 \times 10^5$ | $4.5 \times 10^5$ | Tanaka et al.[16] |
| tight-binding | $1.6 \times 10^5$ | $1.6 \times 10^6$ | Tanaka et al.[16] |
| tight-binding | $2.6 \times 10^5$ | $1.3 \times 10^7$ | Tanaka et al.[16] |
| tight-binding | $4.0 \times 10^5$ | - | Jo et al.[194] |
| first-principles | $0.48 \times 10^5$ | $10^7$ | Guo et al.[17] |
| first-principles | $4.4 \times 10^5$ | $10^{11}$ | Guo et al.[17] |
| first-principles | $3.8 \times 10^5$ | $6 \times 10^6$ | Obstbaum et al.[76] |
| first-principles | $4.56 \times 10^5$ | - | Qiao et al.[193] |

**ACKNOWLEDGEMENT**
The authors acknowledge the support by the Office of Naval Research (N00014-19-1-2143).

**DATA AVAILABILITY STATEMENT**
The data that supports the findings of this study are available within the article.